\documentclass[acmsmall,authorversion,nonacm]{acmart}
\AtBeginDocument{%
  }

\setcopyright{none}

\acmSubmissionID{152}

\begin{document}

\title{The Ephemeral Shadow: Hyperreal Beings in Stimulative Performance}


\author{Dong Zhang}
\email{zhangdong@shanghaitech.edu.cn}
\affiliation{%
  \institution{ShanghaiTech University}
  \city{Shanghai}
  \country{China}
}

\author{Yanjun Zhou}
\email{zhouyj5@shanghaitech.edu.cn}
\affiliation{%
  \institution{ShanghaiTech University}
  \city{Shanghai}
  \country{China}
}

\author{Jingyi Yu}
\authornote{Corresponding author.}
\email{yujingyi@shanghaitech.edu.cn}
\affiliation{%
  \institution{ShanghaiTech University}
  \city{Shanghai}
  \country{China}
}

\renewcommand{\shortauthors}{Zhang et al.}


\begin{abstract}
\textit{The Ephemeral Shadow } is an interactive art installation centered on the concept of "simulacrum," focusing on the reconstruction of subjectivity at the intersection of reality and virtuality. Drawing inspiration from the aesthetic imagery of traditional shadow puppetry, the installation combines robotic performance and digital projection to create a multi-layered visual space, presenting a progressively dematerialized hyperreal experience. By blurring the audience's perception of the boundaries between entity and image, the work employs the replacement of physical presence with imagery as its core technique, critically reflecting on issues of technological subjectivity, affective computing, and ethics. Situated within the context of posthumanism and digital media, the installation prompts viewers to contemplate: as digital technologies increasingly approach and simulate "humanity," how can we reshape identity and perception while safeguarding the core values and ethical principles of human subjectivity?
\end{abstract}



\maketitle

\section{Introduction}
\textit{The Ephemeral Shadow }is an interactive art installation that interrogates the concept of "simulacrum" by delving into the porous boundaries between the real and the virtual, as well as the reconstruction of subjectivity in the digital age. Rooted in Jean Baudrillard's theory of simulacrum\cite{baudrillard1994simulacra}, this work exposes the progressive dislocation of reality through symbolic abstraction, culminating in a state of "hyperreality." Drawing inspiration from the performative traditions of shadow puppetry, \textit{The Ephemeral Shadow }reimagines these aesthetics through robotic performance and digital projection, constructing a space where audiences encounter a hyperreal shadow realm. By situating this work within contemporary discourses on posthumanism and digital mediation, this paper critically examines its symbolic and philosophical dimensions, highlighting its relevance to contemporary society’s ongoing negotiation with technology and identity.

\begin{figure}[t]
    \centering
    \includegraphics[width=0.8\textwidth]{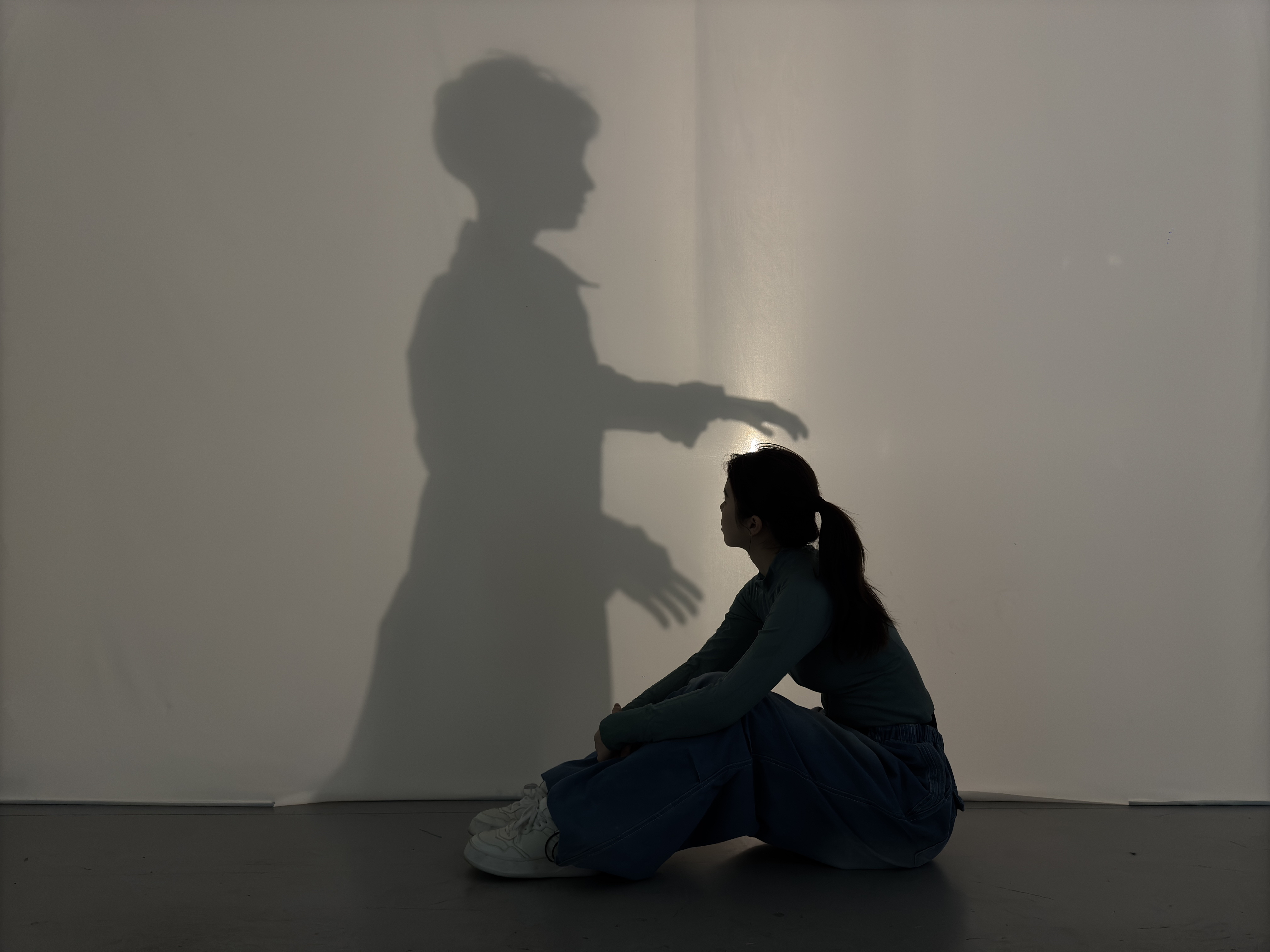}
    \caption{Audience interacting with the projected shadow of \textit{The Ephemeral Shadow}.}
    \label{fig:audience_interaction}
\end{figure}

\section{Motivation}
The creative motivation behind \textit{The Ephemeral Shadow }stems from a profound reflection on the tension between technological realism and audience perception. During early explorations of constructing simulations through traditional technological approaches, we found that relying solely on technologically driven mimicry was insufficient to transcend the uncanny valley and genuinely engage audience empathy. When mechanical replication reaches a high degree of realism, even minor distortions and rigidity amplify the audience's discomfort. This unease arises not only from the stiffness of mechanical performance but also from a deeper fear of disorientation at the blurred boundaries between the virtual and the real.

Sophia, a humanoid robot, serves as a representative case of mechanical simulation. She can mimic expressions such as joy and sadness\cite{zhou2024sophia}. However, this highly realistic replication has significant limitations. While her pre-programmed expressions visually approximate human emotions, they lack the natural fluidity and emotional depth of genuine human interactions, with her "mechanical essence" remaining evident. According to Jean Baudrillard's theory of "preliminary simulation," such realism blurs the line between reality and simulation. Yet, the technological constraints reveal a sense of absence in the original, eliciting discomfort in observers.

To overcome these limitations, we attempted to refine Sophia’s dynamic performance by aligning her movements more closely with human actions through retargeting techniques. By establishing a mapping matrix between BlendShapes and Sophia’s actuators, we enabled real-time facial expression mapping, allowing Sophia to perform nuanced gestures, such as quick blinks or subtle smiles. However, this dynamic enhancement failed to resolve the discomfort entirely. Instead, the closer approximation to "reality" intensified the audience's psychological conflict, exacerbating the uncanny valley effect rather than mitigating it.

To address these technological and psychological challenges, we turned to artistic exploration and philosophical inquiry, seeking solutions that transcend the confines of mechanical mimicry and alleviate audience discomfort. Arising from this line of thought, The Shadow—a robotic image installation—was conceived. By abstracting the mechanical entity and reshaping its interaction design, this work aims to further examine the potential for human–machine coexistence.

\section{The Ephemeral Shadow: Robotic Image Installation}
\textit{The Ephemeral Shadow } is a robotic image installation centered around the "shadow window", inspired by the traditional art form of shadow puppetry and the concept of the "screen self" in internet culture. By projecting Sophia's dynamic performances into two-dimensional images, we aimed to obscure the limitations of her physical entity and abstract her presence into a vibrant and lifelike light-and-shadow interaction.

\begin{figure}[t]
    \centering
    \includegraphics[width=\textwidth]{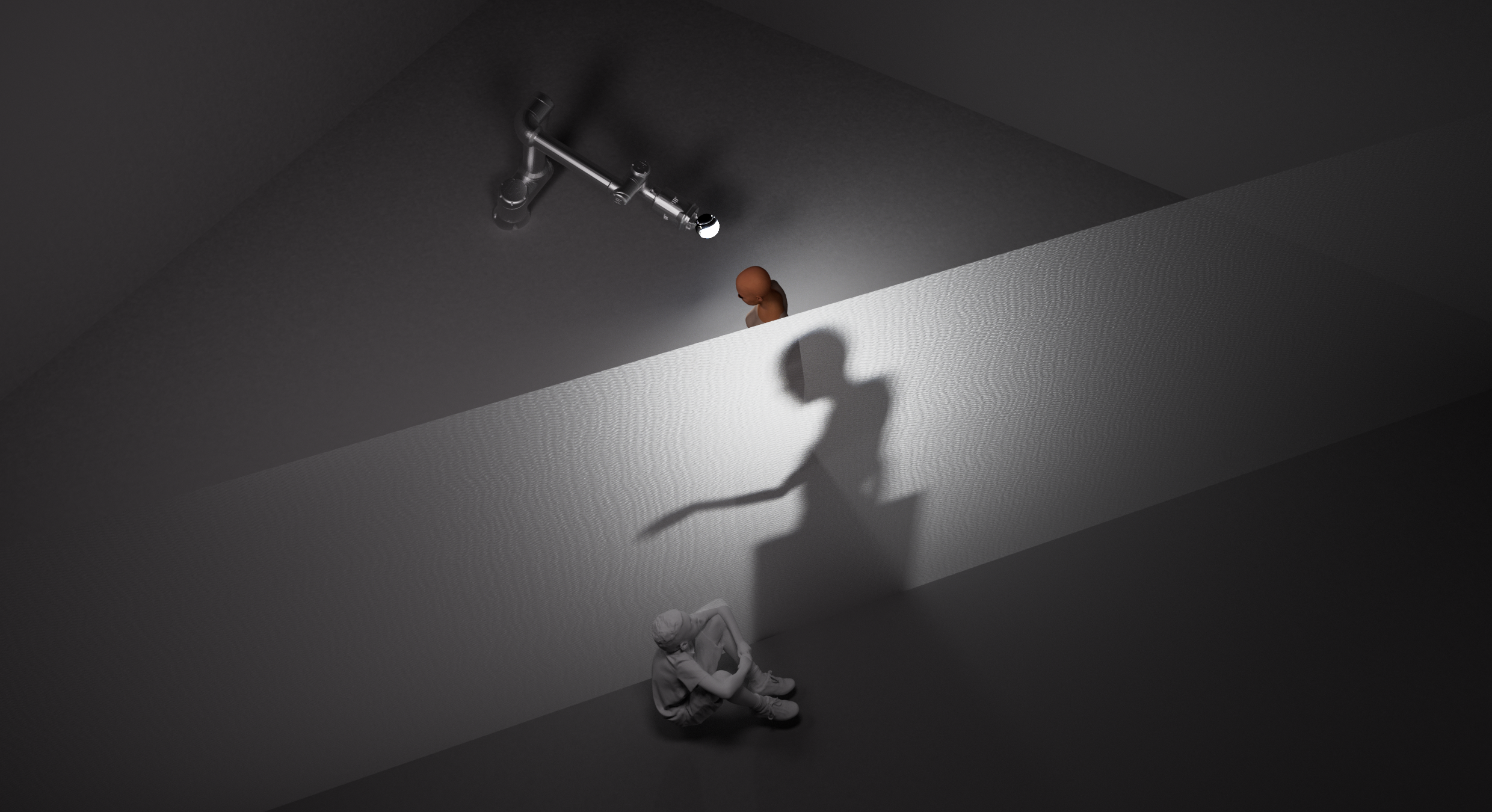}
    \caption{3D rendered schematic of the \textit{The Ephemeral Shadow} installation.}
    \label{fig:ephemeral_shadow_render}
\end{figure}

\subsection{Cultural and Digital Muses: Inspirations of The Shadow"}
Shadow puppetry, as an ancient visual art\cite{ma2023shadow}, projects three-dimensional objects onto a two-dimensional plane through light and shadow, creating a visual experience where the virtual and the real intertwine. The puppeteer remains hidden behind the screen, and the audience can only perceive the narrative through the dynamic movements of the shadows. This form of expression blurs the boundaries between presence and absence, positioning the shadow as the core narrative carrier and emphasizing the aesthetic of "image over entity." 

Additionally, we incorporated Sherry Turkle’s concept of the "screen self" from Life on the Screen\cite{turkle1995life}, exploring how digital media reconstructs individual subjectivity through symbolic projection. Turkle suggests that identity expression in virtual environments not only liberates the self from traditional constraints but also blurs the boundaries between the real and the virtual. In the installation, Sophia’s image is symbolized into an independent entity through the "shadow window", with her mechanical body concealed behind the screen. This design transforms the mechanical dynamics into a symbolic interaction of images, reflecting the mechanism of virtual identity construction and prompting the audience to rethink the complex relationship between real subjectivity and virtual representation in the context of digital media.

Similarly, Zhang Yimou's cinematic masterpiece Shadow\cite{zhang2018shadow}, as another source of inspiration for our work, employs a minimalist color palette and dynamic interactions of light and shadow to evoke a tension between reality and illusion. The film’s juxtaposition of the "real" body and the "shadow self" not only aligns with the installation's exploration of symbolic abstraction and subjectivity but also highlights the concept of the "shadow" as a surrogate (or "double")—both dependent on and independent of its original. This idea has inspired further reflection on the shadow’s potential for autonomy, resistance, and freedom, expanding the interpretation of identity and agency while providing profound cultural and aesthetic references for the installation.

By merging these inspirations, the installation projects Sophia’s vibrant performances onto the "shadow window" via robotic arm-controlled spotlights, creating a dynamic interaction with the audience. This approach elevates the simulacrum to a higher level, concealing the physical constraints of Sophia’s entity while crafting a "hyperreal" interactive environment. At the same time, the interplay of light, shadow, and real-time interaction provides a platform for exploring questions of technological subjectivity and ethics within posthuman discourse, encouraging viewers to critically engage with the intersections of technology, identity, and mediated existence.

\begin{figure}[t]
    \centering
    \includegraphics[width=\textwidth]{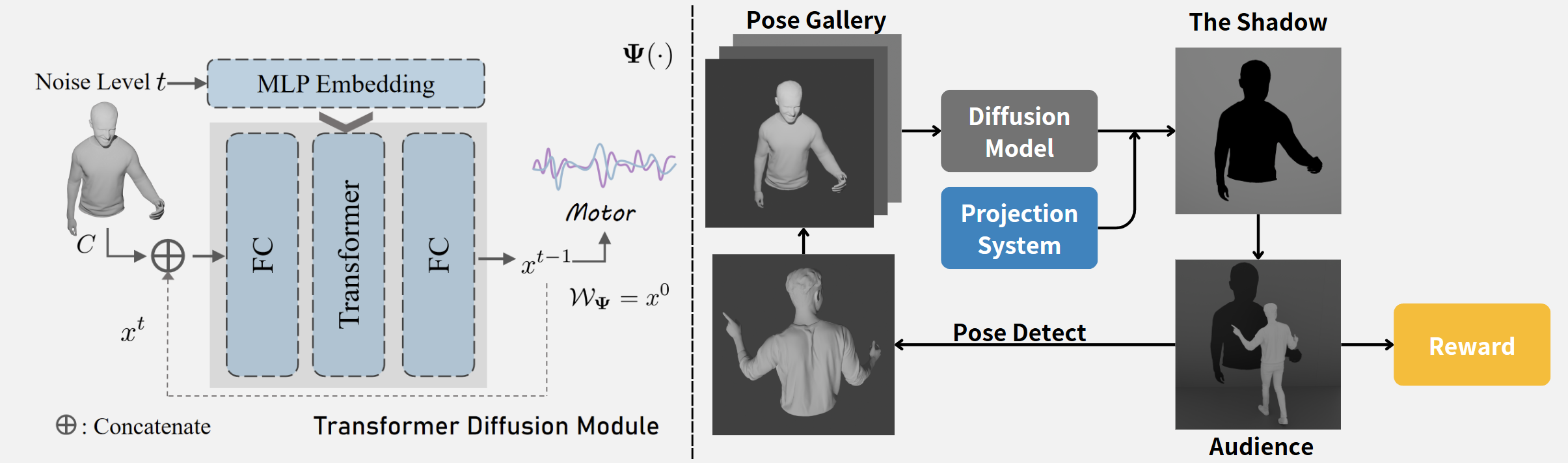}
    \caption{Technical pipeline of \textit{The Ephemeral Shadow}. Left: The denoising network maps human motions and expressions to the robot's motors, enabling action replication and laying the foundation for learning. Right: Diffusion models are trained with reinforcement learning, leveraging audience feedback to enhance interaction and enrich the pose gallery.}
    \label{fig:ephemeral_shadow_pipeline}
\end{figure}

\subsection{Engineering the Simulacrum: A Technical Framework}

In \textit{The Ephemeral Shadow}, Sophia serves as the centerpiece of the installation. Her humanoid expressions are achieved using motor-controlled cords beneath Frubber skin, enabling her to perform complex facial expressions with 33 degrees of freedom (DoF). However, these highly realistic expressions are not directly visible to the audience. Instead, they are projected onto a central white screen using a UR10E robotic arm-controlled spotlight. The light source, equipped with six degrees of freedom, adjusts the angle and intensity of the light, translating Sophia’s movements into two-dimensional images with dynamic variations in size, shape, and contrast. This abstraction enhances Sophia’s presence, allowing viewers to perceive lifelike expressions solely through her shadow, creating a distinctive visual and psychological distance between the real and the simulated.

The projection employs a high-transparency white fabric as the "shadow window" to ensure clarity and depth of the light and shadow interplay. Inspired by the aesthetics of shadow puppetry, this design hides Sophia and the spotlight entirely behind the screen, reinforcing the spatial dichotomy between "backstage" and "stage." This spatial arrangement sharpens the audience's focus on the shadow image while blurring their understanding of "real" and "virtual."

The interactive system is supported by a diffusion model trained on human-robot interaction data. Using an Intel RealSense D435i depth camera, the system captures facial expressions and head movements from performers and audiences, paired with BlendShapes data from Apple ARKit. This enables precise mapping between human actions and robotic motor commands. Initial training data was sourced from performers’ objectless acting, synchronized with shadow puppetry narratives, providing Sophia with story-based templates for her actions.

During live performances, the system employs a reinforcement learning-based diffusion model training strategy, enabling Sophia to continuously optimize her interactive performance by leveraging real-time audience data. The model captures audience feedback behaviors during interactions (such as dwell time and interaction duration) and dynamically adjusts the generated shadow actions. This approach allows Sophia not only to mimic the audience’s real-time expressions and movements but also to generate new motion patterns, progressively evolving into richer and more engaging interactions. By incorporating audience feedback into the learning loop, Sophia’s behavior shifts from simple motion mapping to dynamic generation driven by reinforcement learning. This strategy enables the system to continuously adapt to audience preferences, creating an ever-evolving, highly immersive, multi-layered shadow puppetry performance, ultimately achieving a new level of simulacrum-based performance sophistication.

\begin{figure}[t]
    \centering
    \includegraphics[width=0.8\linewidth]{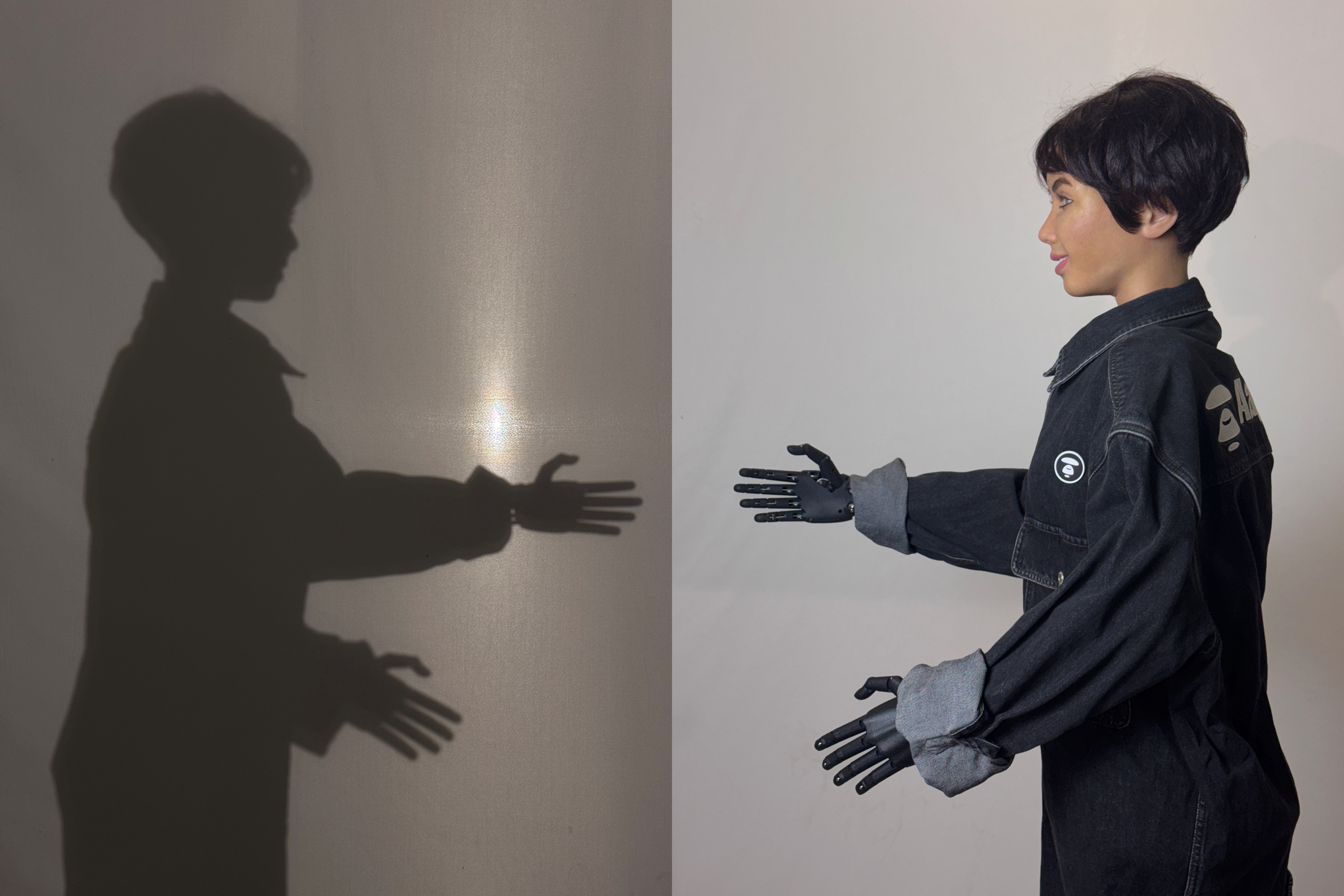}
    \caption{The left side shows the projected shadow image, while the right side depicts the physical entity of Sophia.}
    \label{fig:projection_vs_entity}
\end{figure}

\section{The Performative Simulacrum: Shadows as the New Reality}
At the core of \textit{The Ephemeral Shadow}, the highly realistic Sophia humanoid robot collaborates with a spotlight controlled by a robotic arm to project a dynamically expressive "shadow figure" onto a screen. This image is not Sophia’s true appearance but a secondary representation, abstracted through the interplay of light and motion. The design follows the interactive model of shadow puppetry, "seeing the shadow but not the person," creating an existence that feels both familiar and alien. As Baudrillard's theory of simulacra points out, "It masks the absence of a basic reality." 

Through this setup, \textit{The Ephemeral Shadow } effectively dismantles the binary relationship between "being" and "image." In the absence of direct interaction with Sophia, the robot as an "entity" gradually detaches from its mechanical materiality and gains new identity and meaning through its shadow representation. As viewers watch the shadow, they are compelled to imagine the "reality" behind the screen, but such pursuit remains unfulfilled. The abstraction and ambiguity of the shadow create a sense of disorientation that feels both familiar and strange: it exhibits human-like dynamic traits but lacks concrete physical substance, challenging the audience's perception of the boundary between reality and virtuality. This experience prompts the audience to rethink what "reality" truly means.

This shadow figure, composed of light and motion, amplifies the process of "abstraction": Sophia's physical body is hidden, while its core actions and feedback mechanisms are magnified and expressed through the shadow, directing the audience’s attention to the central interactive mechanism. In this process, the audience becomes both a passive observer of the image and an active constructor of meaning. This viewing approach shifts from the traditional "subject-object" relationship to an interaction model centered on "perception-imagination," where the audience, through observing and speculating on the shadow, participates in the concealment and reconstruction of reality with the installation.
\begin{figure}[t]
    \centering
    \includegraphics[width=0.8\textwidth]{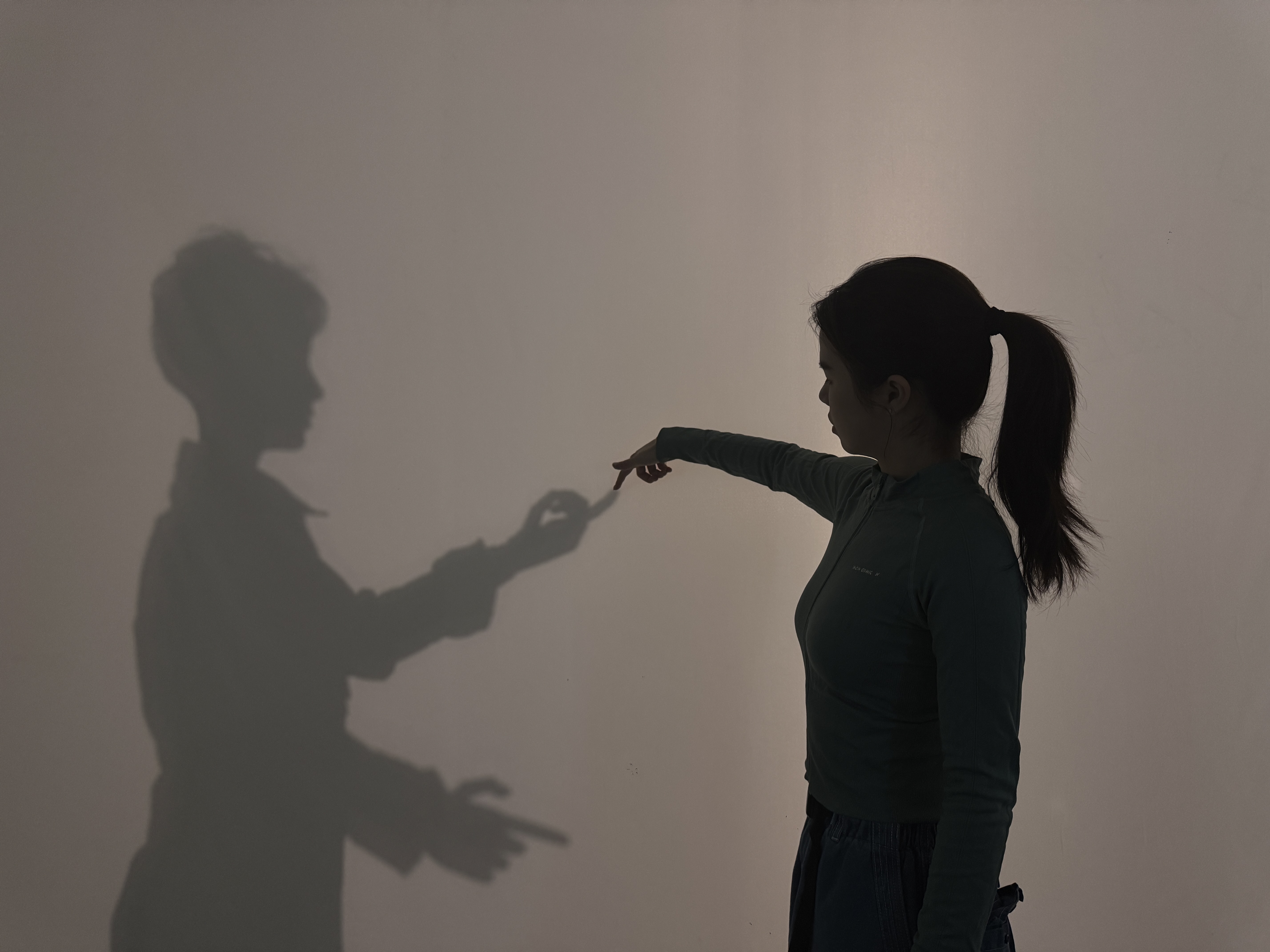}
    \caption{Audience interacting with the projected shadow of \textit{The Ephemeral Shadow}.}
    \label{fig:audience_interaction}
\end{figure}
In this shadow-dominated field, Sophia and its shadow figure no longer represent a simple opposition between "real" and "virtual." Instead, they achieve a dynamic state of existence through the audience’s perception and interaction. What the audience faces is neither a tangible machine nor a purely virtual image, but a composite experience interwoven with technology, art, and philosophical thought. In this experience, viewers might contemplate questions such as: Can an image replace "reality"? Can technology redefine existence? And are the boundaries between humans and machines as clear-cut as traditional concepts suggest? \textit{The Ephemeral Shadow}, through the dual strategy of concealing the entity and presenting the image, blurs traditional boundaries and constructs a new dimension of simulacra. The audience, captivated by the dynamic shadow image, is simultaneously guided to reconsider subjectivity and how technology shapes human perception of reality.

\begin{figure}[t]
    \centering
    \includegraphics[width=\textwidth]{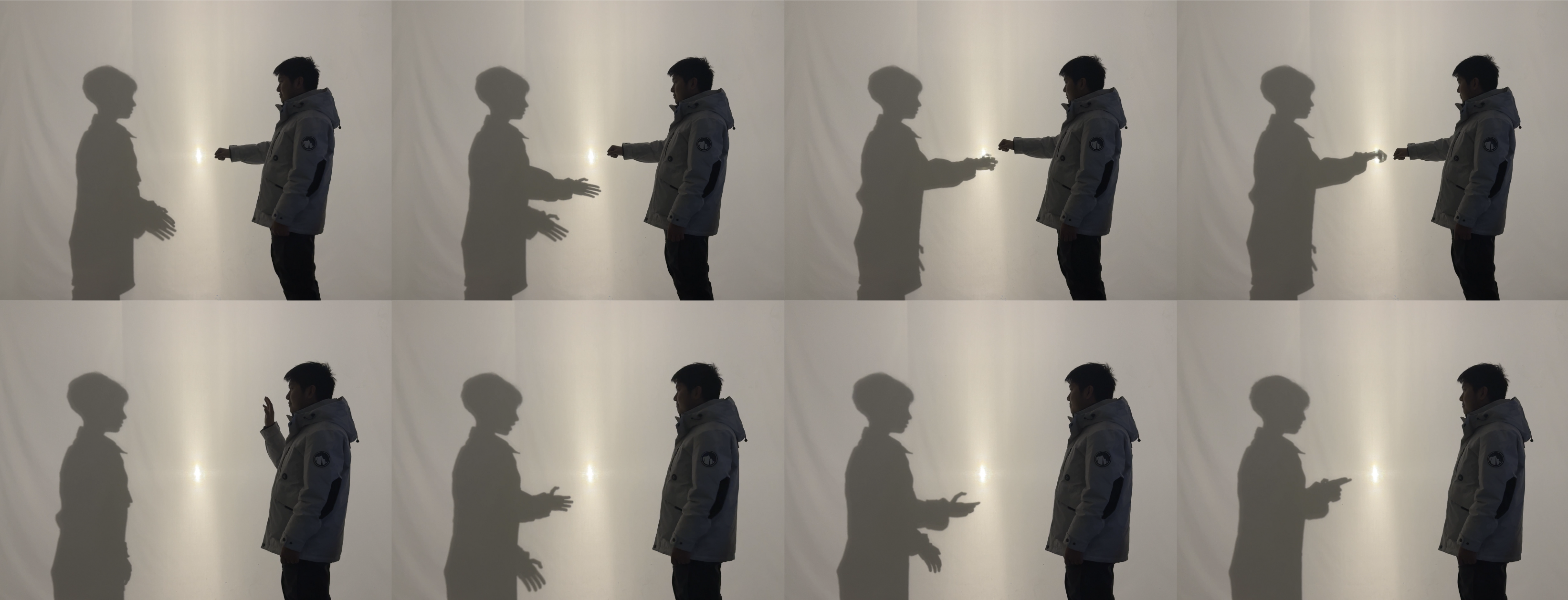}
    \caption{The combined action sequence. The top row illustrates mimicking human fist-bump actions, while the bottom row demonstrates responding with a self-introduction gesture.}
    \label{fig:combined_action_sequence}
\end{figure}

\section{Hyperreality Unveiled: Shadows Beyond Substance}
In \textit{The Ephemeral Shadow}, the design of the "shadow window" progressively dematerializes Sophia’s physical presence into a two-dimensional projection, constructing a “simulation beyond simulation” from the human prototype to the Sophia robot, and finally to the shadow image. This design demonstrates how simulation detaches from its original object and constructs a self-sufficient hyperreal logic.

Sophia's mechanical movements are projected through the "shadow window," where the transformation of light into shadow achieves the "dematerialization of substance." Her bodily actions are simplified into flowing contours of light and shadow, eliminating the rigidity of machinery while enhancing dynamic vividness. This dematerialization avoids the "uncanny valley" effect and imbues the shadow image with an "ethereal humanity." As Baudrillard noted, when a replica becomes more appealing than the original, the original retreats "backstage." In this installation, the "backstage" consists of the Sophia robot and the spotlight system, while the center stage is occupied by the richly layered abstract shadow.

This layered combination of simulation and dematerialization deepens the entanglement between “real” and “virtual.” The audience cannot directly engage with Sophia’s physical presence but interacts with her through the shadow image. This design intentionally blurs the boundaries between reality and virtuality, redirecting the audience’s focus from the physical entity to the performance of the shadow itself. Within this hyperreal atmosphere, the shadow no longer depends on the physical existence of the entity but becomes a form of existence with its own autonomous meaning.

The shadow’s transcendence of the entity is not merely a visual shift but a profound psychological impact. The image in the "shadow window" severs the direct connection between the audience and the physical entity, guiding them into an entirely simulation-constructed order. This “perceptual rupture” not only highlights the image’s dominance in audience perception but also reflects Baudrillard’s description of the fourth-order simulacrum: reality no longer exists in terms of appearance or material form but is entirely reshaped by simulation. This process elevates the audience’s experience beyond material attachment, redirecting their attention to the order created by the shadow image itself. The construction of this hyperreal atmosphere also lays the foundation for the extension of the fourth-order simulacrum, where the shadow image completely escapes the constraints of materiality, becoming an independent and self-sufficient form of existence.

\section{Beyond Simulacra - Creation or Dread?}

In \textit{The Ephemeral Shadow}, we constructed a posthuman evolutionary field that explores the multidimensional aspects of algorithmic autonomy, societal transformation, and ethical challenges through embodied training and feedback learning systems. 

Based on initial data from non-physical performances, Sophia’s model was able to precisely mimic human expressions and actions while continuously optimizing its behavioral outputs by collecting anonymized, real-time audience interaction data. This system not only allowed Sophia to generate emotional actions beyond pre-programmed behaviors, such as hugs, but also expanded the technological agency through algorithm-driven processes. As Katherine Hayles argues\cite{hayles1999posthuman}, robots and humans share deep structural similarities: both achieve sustained evolution through stratified goals and feedback mechanisms. This similarity is evident in their dynamic adaptability to external environments and hierarchical internal structures, enabling self-evolutionary potential.

The amplification of algorithmic bias highlights the concentration of power among technology holders. When developers or institutions influence the learning process through algorithmic settings or data manipulation, technology shifts from being a tool that serves users to becoming a pivotal nexus for the distribution of power and resources. In this process, AI technology begins to act as an "invisible hand of God," reshaping social structures and value orientations in ways that are imperceptible to the public. With both the general public and even policymakers lacking a full understanding of its algorithmic logic and decision-making processes, AI quietly accumulates the power to shape public life, from directing economic trajectories to deeply intervening in social relationships. AI is no longer merely a tool but a self-evolving algorithmic entity, whose invisible control further widens the information and power gap between technology and the public, threatening social fairness and justice.

Blurring the boundaries between technological expression and its underlying mechanisms, this design reveals the pressing issue of insufficient algorithmic transparency. In interacting with the shadow, audiences gradually accept and immerse themselves in the abstraction while remaining unaware of the algorithmic logic and behavioral controls at play. This “black box” effect is not just a technical challenge but a societal issue that concerns fairness and public trust. Whether technology holders can transparently disclose algorithmic logic and decision-making processes remains a central question in the governance of technology.

Sophia’s emotional actions raise further questions: do they genuinely embody human emotions, or are they merely instrumental expressions designed to fulfill pre-programmed objectives? The “designed” nature of these emotional computations challenges the authenticity of human emotions. As affective computing grants technological systems agency, it also redefines the boundaries of emotion. When traditional emotional experiences are regulated or mimicked by algorithms, their authenticity becomes increasingly symbolic. As technology approaches the complexity of emotional expression, its role in the emotional domain shifts from an enabler to a designer. While this evolution strengthens technological agency, it simultaneously undermines traditional notions of anthropocentrism, complicating the balance of emotional authority between humans and machines.

This evolution also signals potential societal transformations. When algorithms not only simulate human behaviors but also guide emotional experiences, human autonomy may be incrementally diminished. As technology transitions from symbolizing emotions to guiding their core experience, cultural authority shifts towards the designers and holders of these systems. The concentration of technological influence could redefine human-technology interactions and impose stricter demands on ethical frameworks.

Through the experimental framework of The Shadow, we observed the potential for autonomous technological evolution and its accompanying sociocultural implications. The installation demonstrates the capability of technology to transcend its instrumental role in emotional domains while revealing how the imbalance of technological power undermines public trust. Only by reinforcing transparency and accountability in algorithmic governance and centering human agency and social ethics in design can technological innovation avoid veering towards concentrated power and ethical alienation.

\begin{figure}[t]
    \centering
    \includegraphics[width=0.8\textwidth]{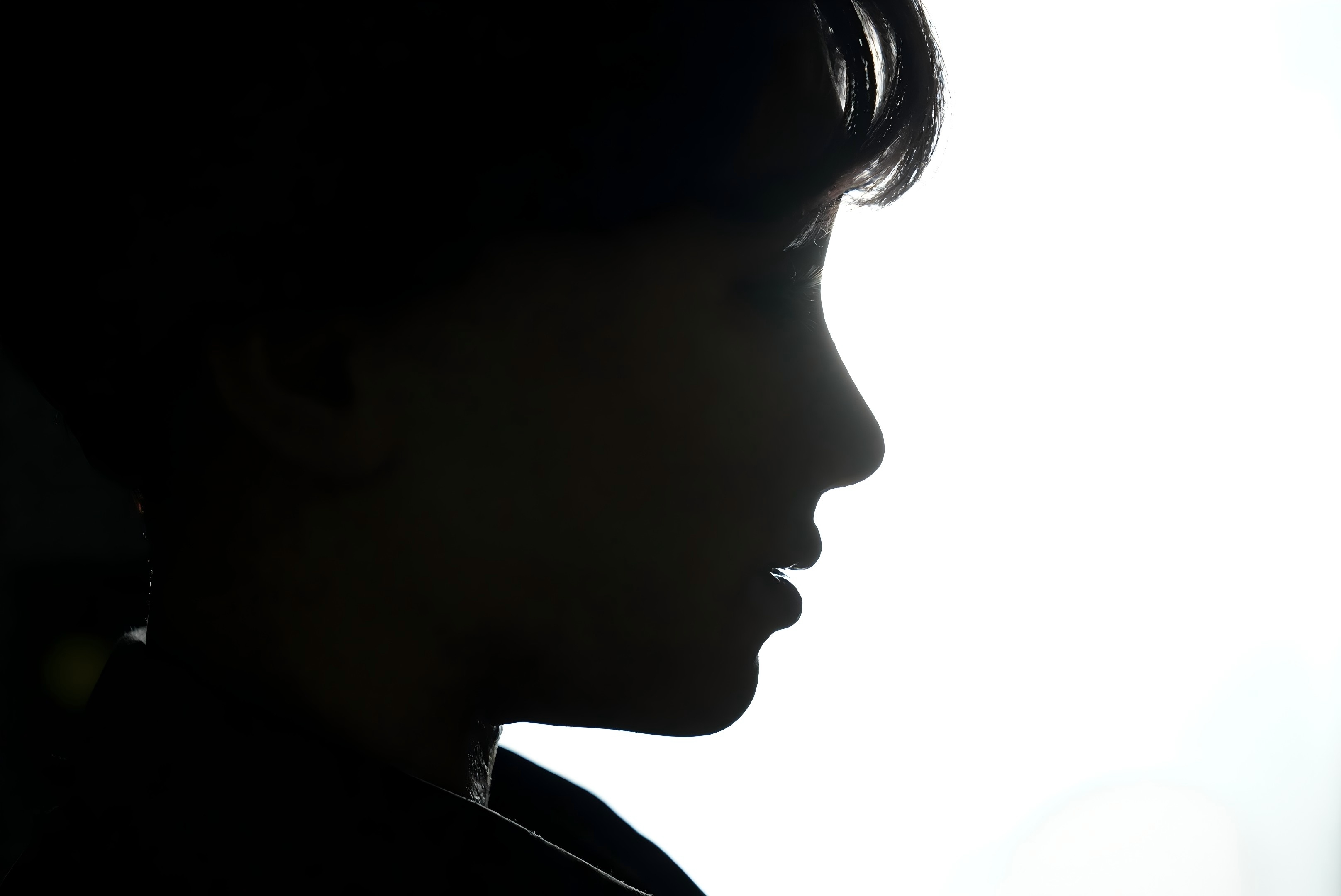} 
    \caption{Through the "shadow window", we gaze into Sophia’s eyes, wondering: in her view, is everything ephemeral?}
    \label{fig:sophia_eye}
\end{figure}

\section{Conclusion}
\textit{The Ephemeral Shadow }symbolizes a new order interwoven with reality and virtuality. Through the construction of multi-layered simulacra and dematerialized design, it reveals the impact of hyperreality on perception and the definition of reality. The core of the work lies in replacing the physical entity with dynamic imagery, immersing the audience in a simulation-centered "hyperreal" realm, demonstrating how images can transcend the entity itself to become a new reality. The multilayered evolution of simulacra showcases the potential for technology to transition from a tool to an autonomous subject, while simultaneously raising profound challenges related to subjectivity, ethics, and societal values.

In conclusion, \textit{The Ephemeral Shadow } calls on us to examine how the boundaries between virtuality and reality are being reconstructed as we embrace the creativity of technology. It urges us to uphold the core values of human subjectivity and ethical responsibility in the design of technology.

\begin{acks}
To Robert, for the bagels and explaining CMYK and color spaces.
\end{acks}

\bibliographystyle{ACM-Reference-Format}
\bibliography{reference}

\nocite{picard1997affective}
\nocite{bostrom2014superintelligence}
\nocite{hansen2006bodies}
\nocite{russell2021artificial}
\nocite{winner1986whale}
\nocite{ellul1954technological}
\nocite{zuboff2018surveillance}
\nocite{christian2011most}
\nocite{manovich2001language}
\nocite{mcluhan1964understanding}
\nocite{egohead2021}
\nocite{trainingdiffusion2023}


\end{document}